\documentclass[aip,apm,reprint,superscriptaddress, showpacs]{revtex4-1}

\usepackage{graphicx}
\usepackage{amsmath}
\usepackage{amssymb}

\begin{document}

\title{Temperature controlled motion of an antiferromagnet-ferromagnet interface within a dopant-graded FeRh epilayer}

\author{C.~Le~Gra\"{e}t}
\affiliation{School of Physics and Astronomy, University of Leeds, Leeds LS2 9JT, United Kingdom}
\affiliation{Laboratoire de Magn\'{e}tisme de Bretagne, 6 avenue Le Gorgeu - CS93837, 29238 Brest Cedex 3, France}

\author{T.~R.~Charlton}
\affiliation{ISIS, Harwell Science and Innovation Campus, Science and Technology
Facilities Council, Rutherford Appleton Laboratory, Didcot, Oxon. OX11
0QX, United Kingdom}

\author{M.~McLaren}
\affiliation{Institute for Materials Research, University of Leeds, Leeds LS2 9JT, United Kingdom}

\author{M.~Loving}
\affiliation{Department of Chemical Engineering, Northeastern University, Boston, MA 02115, USA}

\author{S.~A.~Morley}
\affiliation{School of Physics and Astronomy, University of Leeds, Leeds LS2 9JT, United Kingdom}

\author{C.~J.~Kinane}
\affiliation{ISIS, Harwell Science and Innovation Campus, Science and Technology
Facilities Council, Rutherford Appleton Laboratory, Didcot, Oxon. OX11
0QX, United Kingdom}

\author{R.~M.~D.~Brydson}
\affiliation{Institute for Materials Research, University of Leeds, Leeds LS2 9JT, United Kingdom}

\author{L.~H.~Lewis}
\affiliation{Department of Chemical Engineering, Northeastern University, Boston, MA 02115, USA}

\author{S.~Langridge}
\email{sean.langridge@stfc.ac.uk}
\affiliation{ISIS, Harwell Science and Innovation Campus, Science and Technology
Facilities Council, Rutherford Appleton Laboratory, Didcot, Oxon. OX11
0QX, United Kingdom}

\author{C.~H.~Marrows}
\email{c.h.marrows@leeds.ac.uk}
\affiliation{School of Physics and Astronomy, University of Leeds, Leeds LS2 9JT, United Kingdom}

\begin{abstract}
  Chemically ordered B2 FeRh exhibits a remarkable antiferromagnetic-ferromagnetic phase transition that  is first order. It thus shows phase coexistence, usually by proceeding though nucleation at random defect sites followed by propagation of phase boundary domain walls. The transition occurs at a temperature that can be varied by doping other metals onto the Rh site. We have taken advantage of this to yield control over the transition process by preparing an epilayer with oppositely directed doping gradients of Pd and Ir throughout its height, yielding a gradual transition that occurs between 350~K and 500~K. As the sample is heated, a horizontal antiferromagnetic-ferromagnetic phase boundary domain wall moves gradually up through the layer, its position controlled by the temperature. This mobile magnetic domain wall affects the magnetisation and resistivity of the layer in a way that can be controlled, and hence exploited, for novel device applications.
\end{abstract}


\date{\today}

\maketitle

The chemically ordered alloy FeRh has the B2 structure and has long been known to undergo a magnetostructural phase transition from an antiferromagnetic (AF) to a ferromagnetic (FM) state at a transition temperature $T_\mathrm{t} \sim 380$~K.\cite{Fallot:38,Kouvel:62} Studies of bulk samples have revealed that the transition is accompanied by an isotropic 1\% volume expansion,\cite{deBergevin:61,Zakharov:64} a significant reduction in resistivity,\cite{Kouvel:62} and a substantial release of entropy.\cite{Annaorazov:96} The metamagnetic transition temperature is highly sensitive to the composition $x$ in Fe$_x$Rh$_{1-x}$\cite{Shirane:63,Driel:99} and to chemical doping.\cite{Walter:64,Schinkel:74,Barua:13} $T_\mathrm{t}$ decreases with increased applied magnetic field\cite{Baranov:95} and increases with the application of pressure.\cite{Kamenev:97} Thin film samples of FeRh also undergo such a transition,\cite{Lommel:67,Ohtani:93,Driel:99,Maat:05,Stamm:08,Sharma:11,deVries:13,Bordel2012} and are of interest since they may find technological application in the fields of heat-assisted magnetic recording media\cite{Thiele:03} or resistive memory cells.\cite{Marti:14} Thus, the study of such films is a very active area of research at present, with the aim of developing films suitable for specific applications by growing epitaxially on a variety of substrates and in heterostructures.\cite{Maat:05,Ding:08,Fan:10,Loving:12,Baldasseroni:14,Suzuki:14,Cherifi:14}

The magnetostructural phase transition is well-known to be of first order in this material, and as a result shows hysteresis and phase coexistence. The coexistence of the two phases has been inferred from transport measurements,\cite{Sharma:11} x-ray diffraction experiments,\cite{Kim:09,Loving:2013,mariager:12,deVries:14} x-ray magnetic circular dichroism spectroscopy,\cite{Radu:10} and directly imaged using photoemission electron microscopy.\cite{Baldasseroni:12,Mariager:13,Kinane:14,Baldasseroni:14,Cherifi:14} The co-existing phases are separated by phase-boundary domain walls. In a nominally homogeneous sample these domain walls form around randomly nucleated regions of the different phases. Nevertheless, the very different physical properties of the two different phases suggest that bringing the creation and motion of these phase boundary domain walls under control could be exploited in novel nanomagnetic or spintronic devices. Here we show how a single phase boundary domain wall (DW) may be nucleated and moved in a controllable manner by preparing a film with a vertical gradient of chemical doping density.

\begin{figure}[b]
  \includegraphics[width=6cm]{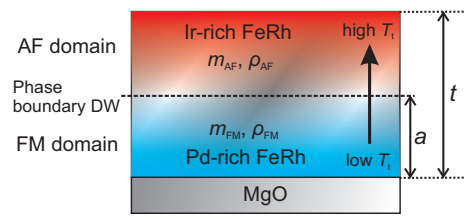}
  \caption{(Color online) Doping gradient concept. The FeRh epilayer has opposing doping gradients of Pd and Ir such that there is a continuous variation of the local $T_\mathrm{t}$ throughout the height of the layer. Thus a phase boundary domain wall exists at the point where the local $T_\mathrm{t} = T$, the experimental temperature, dividing the layer into AF and FM slabs with different magnetization $m$ and resistivity $\rho$. The film is of total thickness $t$ and the FM region is of thickness $a$.  \label{stack}}
\end{figure}

It is well known that doping onto the Rh site can produce large changes in $T_\mathrm{t}$, related to the modification of the density of ($s + d$) valence electrons (see Ref. \onlinecite{Barua:13} and references therein). Here we select two dopants that are well-known to modify $T_\mathrm{t}$ in opposite directions: Pd, which suppresses $T_\mathrm{t}$,\cite{Kushwaha:09} and Ir which enhances it.\cite{Yuasa:94} The sample design is such that the layer is Pd-rich at the bottom and Ir-rich at the top, and thus has a gradient of $T_\mathrm{t}$ throughout its height. At intermediate temperatures the sample will phase separate into two slabs of different magnetic order, AF or FM. We define the co-ordinate normal to the sample plane as $z$. The sample is of thickness $t$ and the phase boundary domain will be at a height $a$, such that for $z<a$ the material is anticipated to be FM and for $z>a$ AF. As $T$ is varied, the phase boundary domain wall can be expected to move up and down, tracking the gradient in $T_\mathrm{t}$. The concept is illustrated in Fig. \ref{stack}.

The growth of our doping gradient film was based on the recipe presented in Ref. \onlinecite{LeGraet@2013}, the main features of which are that the layer is sputtered onto an MgO single crystal substrate held at high temperature (in this case 600~$^\circ$C) at a low rate (of order 0.1~\AA/s) from an alloy target using conventional dc magnetron deposition. In this case a co-sputtering method was used, with targets containing a 3~per~cent doping of Pd and 4~per~cent doping of Ir by atomic number. Thin (sub-nm) layers, deposited sequentially from these two targets, were interleaved to form the desired doping profile. The high substrate temperature means that the dopants can be expected to diffuse, forming a smooth density gradient. The nominal total thickness was 500~\AA. A thin ($\sim 50$~\AA) Al cap was deposited before breaking vacuum once the sample had cooled.

\begin{figure}
  \includegraphics[width=7.5cm]{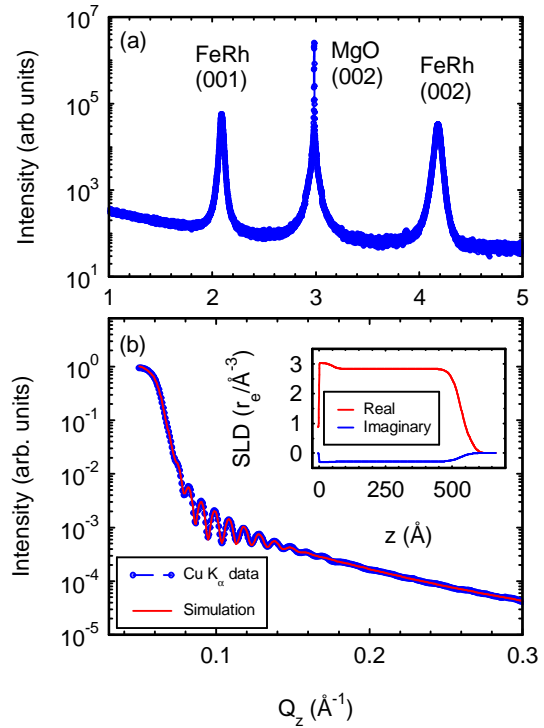}
  \caption{(Color online) X-ray analysis. (a) X-ray diffraction spectrum, with indexed Bragg reflections. The $(001)$ peak for the FeRh is pronounced, indicating B2 ordering. (b) X-ray reflectometry data with fit. The SLD depth profile corresponding to the fit is displayed as an inset, where $r_\mathrm{e} = 2.818 \times 10^5$~\AA\ is the classical electron radius. \label{xrays}}
\end{figure}

The sample was characterized using x-ray diffraction and reflectometry in a conventional Bragg-Brentano geometry using Cu $K_\alpha$ radiation. The x-ray diffraction results are shown in Fig. \ref{xrays}(a), plotted against the out-of-plane component of the wavevector transfer, $Q_z$. It is clear that the layer has grown epitaxially on the MgO substrate. An analysis of the relative integrated intensities of the $(001)$ and $(002)$ Bragg reflections for the FeRh\cite{warren_x-ray_1969,LeGraet@2013} indicates that the degree of chemical order of the B2 structure is $S \approx 0.8$ for this sample. The x-ray reflectometry spectrum is shown in Fig. \ref{xrays}(b). Clear Kiessig fringes can be seen, and the reflectivity persists to large values of $Q_z$ indicating a flat, smooth film. The data were fitted using the GenX code,\cite{genx} returning a total FeRh alloy layer thickness of $t = 525 \pm 6$~\AA.\footnote{The full set of fitting parameters are available in the Supplementary Material, available at [insert URL]} The scattering length density (SLD) depth profile corresponding to the best fit model is displayed as an inset to Fig \ref{xrays}(b).

\begin{figure}
  \includegraphics[width=7.5cm]{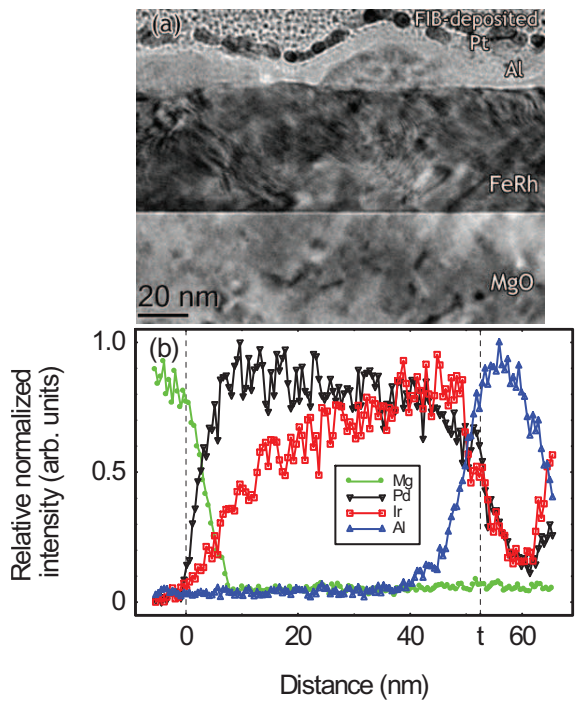}
  \caption{(Color online) TEM characterization. (a) Typical example of the film layout as seen using bright field TEM. (b) Element specific profiles of the heterostructure obtained using STEM-EDX. Mg and Al profiles indicate the substrate and cap respectively. The gradient between Ir and Pd-rich FeRh is visible and fits with the expected result of a smooth gradient. \label{fig:tem}}
\end{figure}

The sample was prepared into a cross-section using a focused ion beam method with a Ga ion source. A layer of Pt was deposited to protect the film from damage. Transmission electron microscope (TEM) characterization was performed using an FEI Tecnai TF20 operating at 200~kV. Fig. \ref{fig:tem}(a) shows a typical example of a bright field image of the FeRh film. The interface it makes with the MgO substrate is sharp while the surface of the FeRh is seen to be smooth as well, although the top surface of the Al cap is rather wavy. Scanning TEM-energy dispersive x-ray spectroscopy (STEM-EDX) line profiles measuring the relative intensity change of the doping elements and the adjacent layers through the height of the film are shown in Fig. \ref{fig:tem}(b) and confirm the presence of a gradient between Ir and Pd-rich within the FeRh. The Fe and Rh profiles are flat within this region (see Fig. S1 in the supplementary information). The TEM probe size was determined to be $\sim 5$~nm from the decay of intensity across the sharp MgO/FeRh interface.

\begin{figure}
  \includegraphics[width=7.5cm]{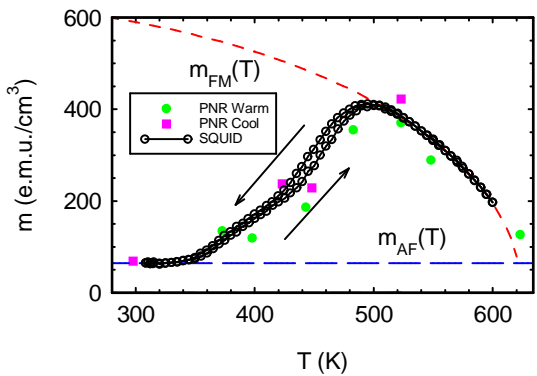}
  \caption{(Color online) Temperature dependence of the magnetization measured at a field of 5~kOe. The open circles represent the data measured using SQUID magnetometry. The long-dash line is a fit to the constant moment $m_\mathrm{AF}$ in the low temperature phase, whilst the short-dash line is a fit of Eq. \ref{eq:moment} to the data above the phase transition. The solid circles and squares indicate the magnetization determined by PNR integrating over the layer thickness. The arrows indicate the direction in which the hysteresis loop is traversed.  \label{squid}}
\end{figure}

The temperature dependence of the magnetization $m$ of the sample was measured using superconducting quantum interference device (SQUID) magnetometry. These $m(T)$ data, measured at a field of 5~kOe, are shown in Fig. \ref{squid}. (This field is enough to suppress the transition temperature by only about 4~K,\cite{Maat:05} a negligible amount in the context of this experiment.) As expected, there is a broad transition that spans the range from $\sim 340$~K to $\sim 500$~K, typical transition temperatures for Pd and Ir doping respectively. The kink at about 440~K indicates a slight discontinuity in the doping gradient. There is some retained ferromagnetism at low temperatures, which we can associate with the top and bottom surfaces of the sample.\cite{Fan:10} This provides a magnetization in the nominally AF state $m_\mathrm{AF} \approx 65$~e.m.u./cm$^3$, which for simplicity we assume to be temperature-independent, represented by the long-dashed line in Fig. \ref{squid}.

The magnetization in the FM phase ($T \gtrsim 500$~K) was fitted with the phenomenological formula\cite{Aharoni:81}
\begin{eqnarray}
m_\mathrm{FM}(T) = 2 m_0 \left[ a^2 + \frac{1-2a}{1-2b} \left( 1-\frac{T}{T_0} \right) \times \right. \nonumber \\
\left. \left( 1 - 2b - \frac{T}{T_0} \right)^c \right]^\frac{1}{2},
\label{eq:moment}
\end{eqnarray}
where $c = 1 + \left[4 S \left( S + 1 \right) \right]^{-1}$ and $S = \frac{1}{2}$. The fit returned the values $m_0 = 640$~e.m.u./cm$^3$, $a = 0.50$, $b = 0.56$, and $T_0 = 57$~K. This expression can then be extrapolated across the whole measurement temperature range, as shown by the short-dashed line in Fig. \ref{squid}, to give an estimate of what the magnetisation would be at each temperature if the material did not undergo its phase transition. The actual data vary smoothly between these two limits as the gradual phase transition takes place, with modest hysteresis of up to about 10~K.

\begin{figure}
  \includegraphics[width=7.5cm]{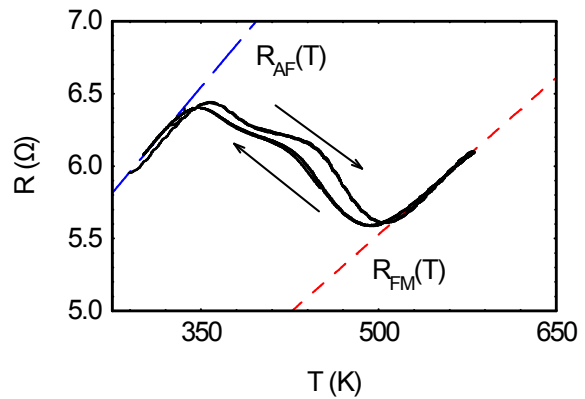}
  \caption{(Color online) Temperature dependence of the resistance measured at zero field. The circles represent the measured data points. The long-dash line is a linear fit to the trend $R_\mathrm{AF}(T)$ in the low temperature phase, whilst the short-dash line is a similar fit to the data above the phase transition to yield $R_\mathrm{FM}(T)$. The arrows indicate the direction in which the hysteresis loop is traversed.  \label{resistivity}}
\end{figure}

As well has having different magnetizations the two phases have very different resistivities,\cite{Kouvel:62,deVries:13} and so the phase separation can also be tracked using resistance measurements. The temperature dependence of the sample resistance $R(T)$, as measured by four point probe, and at zero field, is plotted in Fig. \ref{resistivity}. The resistance is found to depend linearly on temperature deep in the AF and FM phases, and so it is straightforward to fit straight lines $R(T) = aT + r$ to these data and then extrapolate over the whole temperature range. The results of doing so are shown in Fig. \ref{resistivity}, where in the AF phase the fit yields the long-dashed line described by the parameters $a_\mathrm{AF} = 9.74$~m$\Omega$/K and $r_\mathrm{AF} = 3.14~\Omega$, whilst in the FM phase the fit provides the short-dashed line given by $a_\mathrm{AF} = 7.21$~m$\Omega$/K and $r_\mathrm{AF} = 1.92~\Omega$. Again the actual data for $R(T)$ vary smoothly between these two limits.

\begin{figure}
  \includegraphics[width=7.5cm]{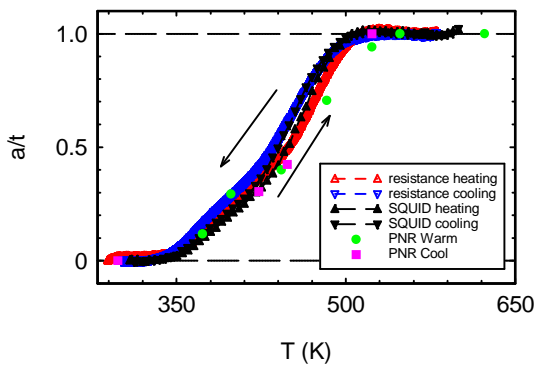}
  \caption{(Color online) Temperature dependence of the phase boundary DW as determined from the SQUID data in Fig. \ref{squid} (using Eq. \ref{eq:magpos}), resistance measurements (using Eq. \ref{eq:respos}) and PNR. The arrows indicate the direction in which the hysteresis loop is traversed. \label{wallpos}}
\end{figure}

The position of the phase boundary DW separating the AF and FM regions can be inferred from these data. In order to do so from the data for $m(T)$, we assume that the material below the phase boundary DW is fully magnetised with $m = m_\mathrm{FM}(T)$ and the material above that point has $m = 0$. Thus we may write
\begin{equation}
a(T) = t  \left( \frac{m(T) - m_\mathrm{AF}(T)}{m_\mathrm{FM}(T) - m_\mathrm{AF}(T)} \right),
\label{eq:magpos}
\end{equation}
where $m_\mathrm{AF}$ is used as an offset. This yields the data for phase boundary DW position $a$ shown as solid triangles in Fig. \ref{wallpos}.

Similarly, by assuming that the sample breaks into two slabs of differing resistivity that conduct the in-plane current in parallel, it is possible to estimate the wall position $a$ from the data for $R(T)$, using the expression
\begin{equation}
a(T) = t \frac{  \left( 1 - \frac{R(T)}{R_\mathrm{AF}(T)} \right)} {\left( \frac{R(T)}{R_\mathrm{FM}(T)} - \frac{R(T)}{R_\mathrm{AF}(T)} \right)},
\label{eq:respos}
\end{equation}
where $R_\mathrm{FM}(T)$ is the resistance of the sample in the fully AF state and $R_\mathrm{FM}(T)$ that in the fully FM state. The results of this procedure are also plotted in Fig. \ref{wallpos} using open triangles. There is a close agreement with the SQUID-derived data, particularly at higher temperatures.

\begin{figure}
  \includegraphics[width=8.5cm]{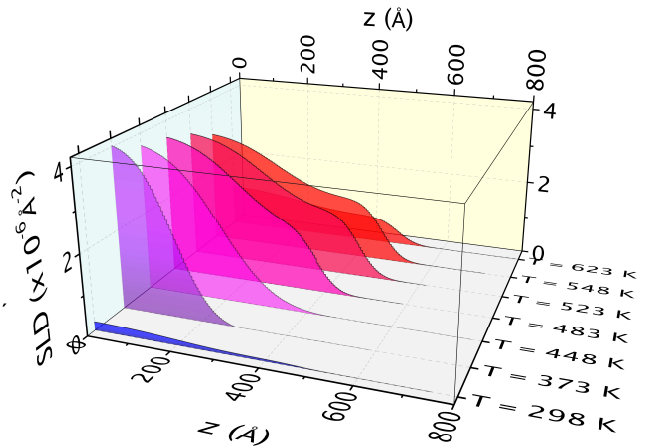}
  \caption{(Color online) Magnetic SLD profile of the layer fitted to the PNR data at various temperatures during a warming process.  \label{pnr}}
\end{figure}

Nevertheless, the measurements of $m(T)$ and $R(T)$ average over the sample and lack spatial resolution. In order to check our assumption that the sample really has phase separated into discrete slabs, we performed polarized neutron reflectometry (PNR) at the PolRef instrument at ISIS. Data obtained from this technique can be fitted to determine the magnetic depth profile of a thin film or multilayer system.\cite{Marrows:09} Our sample was measured under an applied field of 8~kOe (expected to depress $T_\mathrm{t}$ by only $\sim 6$~K) at a series of increasing temperatures.\footnote{The PNR spectra and the full set of fitting parameters are available in the Supplementary Material, available at [insert URL]} These data were fitted with the Refl1D code\cite{Kienzle2011}, constraining the structural parameters to be those returned by the fit to the x-ray reflectometry, and the fitted magnetic SLD at each temperature during a warming cycle is shown in Fig. \ref{pnr}. We see that, as expected, the sample is essentially non-magnetic at the lowest measured temperature of 298~K (but for the small interface-induced magnetism close to the substrate that has been previously observed\cite{Fan:10}), and that as the temperature is increased magnetization appears at the bottom substrate-epilayer interface and gradually spreads upwards through the film as $T$ is raised. The position of the wall obtained from the fits is plotted as solid circles for this warming process in Fig. \ref{wallpos}. The match to the DW position inferred from the SQUID and resistance measurements is excellent. At the highest temperatures the layer starts to depolarize preferentially starting at the film-substrate interface and is consistent with the SQUID measurements. The fitted magnetic SLD can be integrated to return the average magnetization in the sample, which is plotted as solid circles in Fig. \ref{squid}. Again, the PNR fits correspond closely to the magnetization measured by SQUID. Thus, the PNR experiments quantitatively confirm the picture of a phase boundary DW moving up through the layer during heating, controlled by the doping gradient.

AF/FM boundaries within a single material were studied by Saerback \textit{et al.}, where ordered and disordered phases of FePt$_3$, which is AF when taking up the L$1_2$ structure, but ferromagnetic when antisite disorder is present.\cite{Saerbeck:10} Growing epilayers where the deposition temperature is modulated leads to a fully epitaxial structure with modulated antisite defect density and thus a stratification into AF and FM regions. However, the AF/FM domain walls in that case are tied to the boundaries between the layers of different degrees of chemical order, and are not mobile, as in the present case. This mobility could be exploited with more complex inhomogeneous doping profiles, perhaps in two or three dimensions, to yield fine control over the phase transition in terms of both the range of temperatures and distribution in space of the transforming regions. This could be taken advantage of in a wide range of nanomagnetic and--due to the coupling to the electrical resistivity--spintronic applications.

\begin{acknowledgments}
This work was supported by a linked grant through the Materials World Network scheme by the National Science Foundation under Grant No. DMR-0908767 and the UK Engineering and Physical Sciences Research Council, Grant No. EP/G065640/1. Experiments at the ISIS Pulsed Neutron and Muon Source were supported by a beamtime allocation from the Science and Technology Facilities Council. We would like to thank D. Heiman for the use of the SQUID magnetometer.
\end{acknowledgments}

\bibliography{my-bibliography}

\end{document}